\documentclass{ifacconf}

\usepackage{graphicx}      
\usepackage{natbib}        
\defcitealias{WorldEnergyOutlook2021}{IEA, 2021a}
\defcitealias{NetZero2021}{IEA, 2021b}
\usepackage{amsmath,amssymb,amsfonts}
\theoremstyle{plain}

\newtheorem{problem}{Problem}
\newcommand\norm[1]{\left\lVert#1\right\rVert}

\usepackage{algorithmic}
\usepackage{enumitem}
\usepackage{breqn}
\usepackage{mathtools}
\usepackage{url}
\usepackage{multirow}

\usepackage{tikz}
\usepackage{pgfplots}
\pgfplotsset{compat=1.18}

\usetikzlibrary{decorations.pathreplacing,decorations.markings}
\usetikzlibrary{positioning}
\usetikzlibrary{backgrounds,scopes}
\usetikzlibrary {arrows.meta}
\usepackage{adjustbox}

\usepackage{units}
\usepackage{xcolor}

\usepackage[prependcaption,colorinlistoftodos]{todonotes}

\pgfplotsset{every axis/.append style={semithick,tick style={major tick
            length=4pt,semithick,black}}}

\pgfkeys{/pgfplots/x axis shift down/.style={
        x axis line style={yshift=-#1},
        xtick style={yshift=-#1},
        tick align=outside,
        xticklabel shift={#1}}}

\pgfkeys{/pgfplots/y axis shift left/.style={
        y axis line style={xshift=-#1},
        ytick style={xshift=-#1},
        yticklabel shift={#1},
        scaled y ticks = false,
        tick align=outside,
        y tick label style={/pgf/number format/fixed,
        }
    }
}

\pgfplotsset{myPlot/.style={%
        line width = 0.7pt,
        separate axis lines,
        axis x line*=bottom,
        x axis shift down = 3pt,
        enlarge x limits=false,
        axis y line*=left,
        y axis shift left = 6pt,
        enlarge y limits={abs=.25pt},
        enlarge x limits={abs=.25pt},
    }
}

\newcommand{\unitSize}{10mm}
\newcommand{\myScaling}{1.5}

\newcommand{\convGen}[2]{\begin{tikzpicture}[>=latex, line width=#2*0.2mm,scale=#2]
	\draw (-0.5*#1,-0.5*#1) rectangle (0.5*#1,0.5*#1);
	\draw (0,0) circle [radius=0.5*0.8*#1];
	\draw (0,-0.5*0.8*#1) -- (0,0.5*0.8*#1);
\end{tikzpicture}}

\newcommand{\pump}[2]{\begin{tikzpicture}[>=latex, line width=#2*0.2mm,scale=#2]
	\draw (-0.5*#1,-0.5*#1) rectangle (0.5*#1,0.5*#1);	
	\clip (0,0) circle [radius=0.5*0.8*(#1+0.2mm)];
	\draw (0,0) circle [radius=0.5*0.8*#1];
	\draw (-#1,0) -- (0.4*0.8*#1,0.4*0.8*#1);
	\draw (-#1,0) -- (0.4*0.8*#1,-0.4*0.8*#1);
\end{tikzpicture}}

\newcommand{\battery}[2]{\begin{tikzpicture}[>=latex, line width=#2*0.2mm,scale=#2]
	\draw (-0.5*#1,-0.5*#1) rectangle (0.5*#1,0.5*#1);
	\draw (0,-0.5*0.8*#1) -- (0,-0.1*#1) -- (0.5*0.7*#1,-0.1*#1) -- (-0.5*0.7*#1,-0.1*#1);
	\draw (0,0.5*0.8*#1) -- (0,0.1*#1) -- (0.5*0.7*#1,0.1*#1) -- (-0.5*0.7*#1,0.1*#1);	 
\end{tikzpicture}}

\newcommand{\thermalStorage}[2]{\begin{tikzpicture}[>=latex, line width=#2*0.2mm,scale=#2]
	\draw (-0.5*#1,-0.5*#1) rectangle (0.5*#1,0.5*#1);
	\draw (-0.3*0.8*#1,-0.35*0.8*#1) rectangle (0.3*0.8*#1,0.35*0.8*#1);	
	\draw (-0.3*0.8*#1,-0.35*0.8*#1) .. controls (-0.3*0.8*#1,-0.45*#1) and (0.3*0.8*#1,-0.45*#1) .. (0.3*0.8*#1,-0.35*0.8*#1);
	\draw (-0.3*0.8*#1,0.35*0.8*#1) .. controls (-0.3*0.8*#1,0.45*#1) and (0.3*0.8*#1,0.45*#1) .. (0.3*0.8*#1,0.35*0.8*#1);
	\draw [fill=black] (-0.5*0.9*#1,-0.05*#1) -- (-0.4*0.8*#1,0.05*#1) -- (-0.4*0.8*#1,-0.05*#1) -- (-0.5*0.9*#1,0.05*#1) -- (-0.5*0.9*#1,-0.05*#1) -- (-0.4*0.8*#1,0.05*#1);
	\draw (-0.4*0.8*#1,0) -- (-0.3*0.8*#1,0);
	\draw [fill=black] (0.5*0.9*#1,-0.05*#1) -- (0.4*0.8*#1,0.05*#1) -- (0.4*0.8*#1,-0.05*#1) -- (0.5*0.9*#1,0.05*#1) -- (0.5*0.9*#1,-0.05*#1) -- (0.4*0.8*#1,0.05*#1);
	\draw (0.4*0.8*#1,0) -- (0.3*0.8*#1,0);
\end{tikzpicture}}

\newcommand{\heatEx}[2]{\begin{tikzpicture}[>=latex, line width=#2*0.2mm,scale=#2]
	\draw (0,0) circle [radius=0.5*0.8*#1];	
	\draw (-0.5*0.8*#1,-0.6*0.5*0.8*#1) -- (0.6*0.5*0.8*#1,-0.6*0.5*0.8*#1) -- (0,0) -- (0.6*0.5*0.8*#1,0.6*0.5*0.8*#1) -- (-0.5*0.8*#1,0.6*0.5*0.8*#1);
\end{tikzpicture}}

\newcommand{\Solar}[2]{\begin{tikzpicture}[>=latex, line width=#2*0.2mm,scale=#2]
	\draw (-0.5*#1,-0.5*#1) rectangle (0.5*#1,0.5*#1);	
	\draw (-0.3*#1,0.25*#1) [fill=black] -- (0.3*#1,0.25*#1);
	\draw (-0.35*#1,0.1*#1) [fill=black] -- (0.35*#1,0.1*#1);
	\draw (-0.4*#1,-0.05*#1) [fill=black] -- (0.4*#1,-0.05*#1);
	\draw (-0.45*#1,-0.2*#1) [fill=black] -- (0.45*#1,-0.2*#1);
	
	\draw (0.45*#1,-0.2*#1) [fill=black] -- (0.3*#1,0.25*#1);
	\draw (-0.45*#1,-0.2*#1) [fill=black] -- (-0.3*#1,0.25*#1);
	\draw (0.17*#1,-0.2*#1) [fill=black] -- (0.07*#1,0.25*#1);
	\draw (-0.17*#1,-0.2*#1) [fill=black] -- (-0.07*#1,0.25*#1);
	
	\draw (-0.08*#1,-0.2*#1) [fill=black] -- (0.08*#1,-0.2*#1) -- (0.15*#1,-0.45*#1) -- (-0.15*#1,-0.45*#1);
	
	\draw (-0.23*#1,0.25*#1) .. controls  (-0.1,0.43) and (0.1,0.43) .. (0.23,0.25);
	
	\draw (-0.23*#1,0.28*#1) -- (-0.4*#1,0.38*#1);
	\draw (0.23*#1,0.28*#1) -- (0.4*#1,0.38*#1);
	
	\draw (-0.15*#1,0.36*#1) -- (-0.22*#1,0.45*#1);
	\draw (0.15*#1,0.36*#1) -- (0.22*#1,0.45*#1);
	
	\draw (0*#1,0.405*#1) -- (0*#1,0.485*#1);

\end{tikzpicture}}

\newcommand{\electricLoad}[2]{\begin{tikzpicture}[>=latex, line width=#2*0.2mm,scale=#2]
	\draw (-0.5*#1,-0.5*#1) rectangle (0.5*#1,0.5*#1);
	\draw (0.4*#1,-0.4*#1) [fill=black] -- (0.4*#1,0.15*#1);
	\draw (-0.4*#1,-0.4*#1) [fill=black] -- (-0.4*#1,0.15*#1);
	\draw (-0.4*#1,-0.4*#1) [fill=black] -- (0.4*#1,-0.4*#1);
	\draw (-0.4*#1,0.15*#1) [fill=black] -- (0*#1,0.4*#1);
	\draw (0.4*#1,0.15*#1) [fill=black] -- (0*#1,0.4*#1);
	\draw (0,-0.05) circle [radius=0.5*0.5*#1];
	\draw (-0.2*#1,-0.05) .. controls (-0.02*#1,0.4*#1) and (0.02*#1,-0.5*#1) .. (0.2*#1,-0.05);
\end{tikzpicture}}
\definecolor{vgRed}{RGB}{193, 48, 24}
\definecolor{vgDeepBlue}{RGB}{24,118,172}
\definecolor{vgOrange}{RGB}{243, 111, 19}
\definecolor{vgOrangeII}{RGB}{139, 70, 50}
\definecolor{vgYellow}{RGB}{235, 203, 56}
\definecolor{vgGreen}{RGB}{162, 185, 105}
\definecolor{vgLightBlue}{RGB}{13, 149, 188}
\definecolor{vgDarkBlue}{RGB}{6, 56, 81}

\begin{document}
\begin{frontmatter}

\title{A Predictive Operation Controller for an Electro-Thermal Microgrid Utilizing Variable Flow Temperatures}

\thanks[footnoteinfo]{This research is partly supported by the German Federal Government, the Federal Ministry of Education and Research and the State of Brandenburg within the framework of the joint project EIZ: Energy Innovation Center (project number 03SF0693E) with funds from the Structural Development Act for coal-mining regions.
}

\author[First]{Max Rose}
\author[Second]{Christian A. Hans}
\author[First,Third]{Johannes Schiffer}

\address[First]{Fraunhofer Research Institution for Energy Infrastructures and Geothermal Systems IEG,
   03046 Cottbus, Germany (e-mail: \{max.rose, johannes.schiffer\}@ieg.fraunhofer.de).}
\address[Second]{University of Kassel, Wilhelmsh\"oher Allee 71--73, 34121 Kassel, (e-mail: hans@uni-kassel.de)}
\address[Third]{Brandenburg University of Technology Cottbus-Senftenberg, 03046 Cottbus, Germany (e-mail: schiffer@b-tu.de).}

\begin{abstract} 
We propose an optimal operation controller for an electro-thermal microgrid.
Compared to existing work, our approach increases flexibility by operating the thermal network with variable flow temperatures and in that way explicitly exploits its inherent storage capacities.
To this end, the microgrid is represented by a multi-layer network composed of an electrical and a thermal layer.
We show that the system behavior can be represented by a discrete-time state model derived from DC power flow approximations and 1d Euler equations.
Both layers are interconnected via heat pumps.
By combining this model with desired operating objectives and constraints, we obtain a constrained convex optimization problem.
This is used to derive a model predictive control scheme for the optimal operation of electro-thermal microgrids.
The performance of the proposed operation control algorithm is demonstrated in a case study.
\end{abstract}


\end{frontmatter}

\section{Introduction}

Experts around the world concur that CO$_2$-neutrality in energy supply utilizing predominantly renewable energy sources (RES) is key to preserve a livelihood for humanity on the planet \citepalias{WorldEnergyOutlook2021}.
In order to reach climate-neutrality across all energy sectors, an increased RES-based electrification of the heating, cooling and mobility sector is required \citepalias{NetZero2021}. 
This entails the transformation of once independently operated energy sectors to cross-sectoral generation and distribution of electricity, heat and gas, commonly referred to as sector coupling \citep{Ramsebner2021}.


In 2015, according to \cite{trier_guidelines_2018}, about $50\%$ of final energy consumption in $28$ European countries was used for heating purposes. 
Consequently, the sector-coupled operation of electrical and heat grids plays a central role when it comes to climate-neutrality \citep{trier_guidelines_2018}.
This motivates our focus on electro-thermal energy systems in the present paper.

Since most existing energy systems are based on centralized and controllable power generation from storable fossil fuels, major challenges of integrating high shares of RES into existing infrastructure arise from their intermittent nature and decentralized installation \citep{worighi_integrating_2019}.
To overcome these challenges, the overall energy system can be partitioned into subsystems, i.e., microgrids, to then obtain a system of reasonable size which can be controlled more easily \citep{schiffer2016survey}. 
Flexible operating strategies for microgrids that are sector coupled energy systems (SCES) which maintain power balance despite fluctuating RES infeed and utilize synergies of different energy sectors are needed \citep{Ramsebner2021}.
These operating strategies can be formulated as multi-objective optimization problem \citep{martinez_cesena_energy_2019}.
In this context, model predictive control (MPC) is a well-suited optimization-based approach which models the dynamics of SCES, while taking operational constraints explicitly into account \citep{elmouatamid_review_2021}. 

For applying MPC, generally a dynamic model of the system is required \citep{morari_model_1999}.
So far, in SCES operated by MPC, algebraic power balance equations are predominantly used to model the power flows \citep[see e.g.][]{carli_robust_2022}.
Likewise, available optimization-based control strategies for those systems usually employ simplified, purely stationary models of the structure and the dynamics of grids \citep{martinez_cesena_energy_2019}.
In detail, most approaches disregard the inherent dynamics of district heating grids (DHGs) which often represent a main source of flexibility for their operation \citep{vandermeulen_controlling_2018}.


In what follows, we briefly discuss optimization-based control strategies for DHGs utilizing thermal grid dynamics as well as recent modeling approaches of DHGs.
\cite{giraud_optimal_2017} used a modeling approach that describes the delay of temperature transport throughout DHGs for a predictive control strategie of DHGs.
However, this work neither considers a detailed grid model nor a sector-coupled operation.
For control approaches of SCES that include DHGs, a drive towards robust optimization-based control strategies can be observed.
\cite{carli_robust_2022} developed a robust MPC scheme for demand side management, while \cite{martinez_cesena_energy_2019} proposed a robust two-stage optimization scheme for SCES.
Those approaches include a structured handling of energy demand and supply uncertainties, but neglect the inherent flexibility potential of DHGs, since no thermal grid dynamics are modeled.
To model DHGs in great physical detail, \cite{krug_nonlinear_2021} presented a nonlinear optimization model where $1$d Euler equations were used to derive a model of the water flow in a DHG.
On top of that, \cite{machado_modeling_2022} built and described a modeling approach of DHGs, whose passivity properties were shown.
Based on this work, detailed DHG models can be derived.


Motivated by these developments, we derive a discrete-time state model of an electro-thermal microgrid (ETMG) that explicitly accounts for temperature dynamics together with an MPC to unleash the inherent flexibility of the ETMG during operation.
More precisely, the main contributions of the paper are three-fold:
At first, building upon the works of \cite{krug_nonlinear_2021,machado_modeling_2022}, we develop a discrete-time state model describing the storage and thermal dynamics of a DHG.
Then, we combine this DHG model with the modeling approach for electrical microgrids from \cite{hans_operation_2021}.
This yields a coherent discrete-time model of an ETMG.
Second, based on this model, we formulate an optimal operation controller using MPC, which considers both economic as well as efficiency-based operation objectives.
Third, we demonstrate in a case study that the operation controller exploits the storage as well as inherent DHG dynamics as flexibility potential to minimize operating costs and improve the operational efficiency.

The remainder of the paper is structured as follows.
In Section \ref{sec:ModelETMG}, we describe the modeling of ETMGs. 
Then, we derive an MPC approach for the developed model in Section \ref{sec:OptimizationModel}. 
A case study demonstrating the flexibilization potentials is presented in Section \ref{sec:CaseStudy}.
The presented work is summarized and conclusions are drawn in Section \ref{sec:Conclusion}.

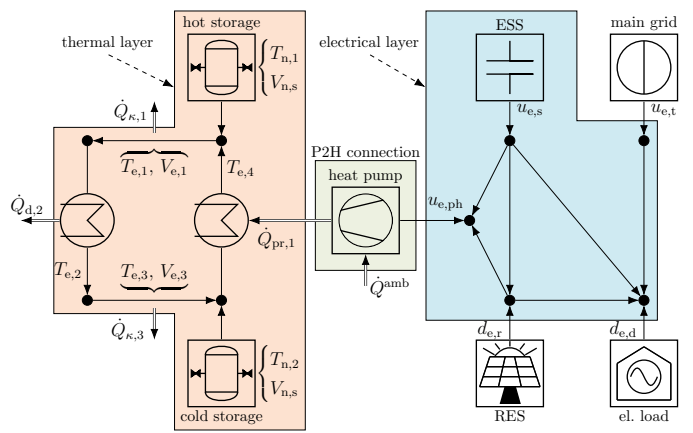
\begin{figure}[!b]
\begin{center}
	\resizebox{\linewidth}{!}{

\begin{tikzpicture}
	[intersec/.style={circle, draw=black, fill=black, minimum size=0.15*\unitSize*\myScaling}, inner sep=0, >=latex]
	\filldraw [fill=vgLightBlue!20] (0.25*\unitSize*\myScaling,3.15*\unitSize*\myScaling) -- 
							(2.5*\unitSize*\myScaling,3.15*\unitSize*\myScaling) --
							(2.5*\unitSize*\myScaling,1.5*\unitSize*\myScaling) --  
							(3.7*\unitSize*\myScaling,1.5*\unitSize*\myScaling) --
							(3.7*\unitSize*\myScaling,-1.5*\unitSize*\myScaling) --
							(0.25*\unitSize*\myScaling,-1.5*\unitSize*\myScaling) -- cycle;
							
	\filldraw [fill=vgOrange!20] (-1.55*\unitSize*\myScaling,3.15*\unitSize*\myScaling) -- 
							(-3.5*\unitSize*\myScaling,3.15*\unitSize*\myScaling) --
							(-3.5*\unitSize*\myScaling,1.4*\unitSize*\myScaling) --  
							(-5.3*\unitSize*\myScaling,1.4*\unitSize*\myScaling) --
							(-5.3*\unitSize*\myScaling,-1.4*\unitSize*\myScaling) --
							(-3.5*\unitSize*\myScaling,-1.4*\unitSize*\myScaling) --
							(-3.5*\unitSize*\myScaling,-3.15*\unitSize*\myScaling) --
							(-1.55*\unitSize*\myScaling,-3.15*\unitSize*\myScaling) --  cycle;
							
	\filldraw [fill=vgGreen!20] (0.1*\unitSize*\myScaling,0.9*\unitSize*\myScaling) -- 
							(-1.4*\unitSize*\myScaling,0.9*\unitSize*\myScaling) -- 
							(-1.4*\unitSize*\myScaling,-0.75*\unitSize*\myScaling) --
							(0.1*\unitSize*\myScaling,-0.75*\unitSize*\myScaling) -- cycle;
							
	\node [label={[label distance=0.04*\unitSize*\myScaling]90:thermal layer},fill=none] (thermalLabel) at (-4.5*\unitSize*\myScaling,2.5*\unitSize*\myScaling) {};
	\node (thermalLabelSink) at (-3.5*\unitSize*\myScaling,2*\unitSize*\myScaling) {};
	\draw[dashed,-{Latex[length=1.5mm*\myScaling, width=1.1mm*\myScaling]}] (thermalLabel) to node[] {} (thermalLabelSink);
	
	\node [label={[label distance=0.0*\unitSize*\myScaling]90:P2H connection},fill=none] (conversionLabel) at (-0.65*\unitSize*\myScaling,1.4) {};
	
	\node [label={[label distance=0.04*\unitSize*\myScaling]90:electrical layer},fill=none] (electricalLabel) at (-0.6*\unitSize*\myScaling,2.5*\unitSize*\myScaling) {};
	\node (electricalLabelSink) at (0.25*\unitSize*\myScaling,2*\unitSize*\myScaling) {};
	\draw[dashed,-{Latex[length=1.5mm*\myScaling, width=1.1mm*\myScaling]}] (electricalLabel) to node[] {} (electricalLabelSink);
		
	\node (ne3) at (0.9*\unitSize*\myScaling,0) [intersec] {};
	\node (ne2) at (1.5*\unitSize*\myScaling,1.2*\unitSize*\myScaling) [intersec] {};
	\node [label={[label distance=0.04*\unitSize*\myScaling]90:ESS}] (battery) at (1.5*\unitSize*\myScaling,2.3*\unitSize*\myScaling) {\battery{\unitSize}{\myScaling}};
	\node (ne1) at (3.5*\unitSize*\myScaling,1.2*\unitSize*\myScaling) [intersec] {};
	\node [label={[label distance=0.04*\unitSize*\myScaling]90:main grid}] (conventional generator) at (3.5*\unitSize*\myScaling,2.3*\unitSize*\myScaling) {\convGen{\unitSize}{\myScaling}};
	\node (ne4) at (1.5*\unitSize*\myScaling,-1.2*\unitSize*\myScaling) [intersec] {};
	\node [label={[label distance=0.04*\unitSize*\myScaling]-90:RES}] (RES) at (1.5*\unitSize*\myScaling,-2.3*\unitSize*\myScaling) {\Solar{\unitSize}{\myScaling}};
	\node (ne5) at (3.5*\unitSize*\myScaling,-1.2*\unitSize*\myScaling) [intersec] {};
	\node [label={[label distance=0.04*\unitSize*\myScaling]-90:el. load}] (electrical load) at (3.5*\unitSize*\myScaling,-2.3*\unitSize*\myScaling) {\electricLoad{\unitSize}{\myScaling}};
	
	\node [label={[label distance=0.04*\unitSize*\myScaling]90:heat pump}] (power to heat) at (-0.65*\unitSize*\myScaling,0) {\pump{\unitSize}{\myScaling}};
	\node [label={[label distance=0.04*\unitSize*\myScaling]0:\large $\dot{Q}^{\mathrm{amb}}$}] (heat resource) at (-0.65*\unitSize*\myScaling,-1*\unitSize*\myScaling) [] {};

	\node (producer) at (-2.8*\unitSize*\myScaling,0) {\heatEx{\unitSize}{\myScaling}};
	\node (nt1) at (-2.8*\unitSize*\myScaling,1.2*\unitSize*\myScaling) [intersec] {};
	\node [label={[label distance=0.04*\unitSize*\myScaling]90:hot storage},label={[label distance=0.0*\unitSize*\myScaling]0:\large $\begin{cases}T_{\mathrm{n,}1} \\ V_{\mathrm{n,s}} \end{cases}$}] (hot storage) at (-2.8*\unitSize*\myScaling,2.3*\unitSize*\myScaling) {\thermalStorage{\unitSize}{\myScaling}};
	\node (nt4) at (-2.8*\unitSize*\myScaling,-1.2*\unitSize*\myScaling) [intersec] {};
	\node [label={[label distance=0.04*\unitSize*\myScaling]-90:cold storage},label={[label distance=0.0*\unitSize*\myScaling]0:\large $\begin{cases}T_{\mathrm{n,}2} \\ V_{\mathrm{n,s}} \end{cases}$}] (cold storage) at (-2.8*\unitSize*\myScaling,-2.3*\unitSize*\myScaling) {\thermalStorage{\unitSize}{\myScaling}};
	\node (consumer) at (-4.8*\unitSize*\myScaling,0) {\heatEx{\unitSize}{\myScaling}};
	\node (heat sink) at (-5.8*\unitSize*\myScaling,0) [] {};
	\node (nt2) at (-4.8*\unitSize*\myScaling,1.2*\unitSize*\myScaling) [intersec] {};
	\node (nt3) at (-4.8*\unitSize*\myScaling,-1.2*\unitSize*\myScaling) [intersec] {};
	\node (hot pipe loss 1) at (-3.8*\unitSize*\myScaling,1.3*\unitSize*\myScaling) [] {};
	\node (hot pipe loss 2) at (-3.8*\unitSize*\myScaling,1.8*\unitSize*\myScaling) [] {};
	\node (cold pipe loss 1) at (-3.8*\unitSize*\myScaling,-1.3*\unitSize*\myScaling) [] {};
	\node (cold pipe loss 2) at (-3.8*\unitSize*\myScaling,-1.8*\unitSize*\myScaling) [] {};

	\draw[-{Latex[length=1.5mm*\myScaling, width=1.1mm*\myScaling]}] (conventional generator) to node[auto,xshift=0.1*\unitSize*\myScaling,yshift=0.12*\unitSize*\myScaling] {\large $u_{\mathrm{e,t}}$} (ne1);
	\draw[-{Latex[length=1.5mm*\myScaling, width=1.1mm*\myScaling]}] (battery) to node[auto,xshift=0.1*\unitSize*\myScaling,yshift=0.12*\unitSize*\myScaling] {\large $u_{\mathrm{e,s}}$} (ne2);
	\draw[-{Latex[length=1.5mm*\myScaling, width=1.1mm*\myScaling]}] (power to heat) to node[auto,yshift=0.15*\unitSize*\myScaling,xshift=0.2*\unitSize*\myScaling] {\large $u_{\mathrm{e,ph}}$} (ne3);
	\draw[-{Latex[length=1.5mm*\myScaling, width=1.1mm*\myScaling]}] (RES) to node[auto,xshift=-0.1*\unitSize*\myScaling,yshift=-0.12*\unitSize*\myScaling] {\large $d_{\mathrm{e,r}}$} (ne4);
	\draw[-{Latex[length=1.5mm*\myScaling, width=1.1mm*\myScaling]}] (electrical load) to node[auto,xshift=-0.1*\unitSize*\myScaling,yshift=-0.12*\unitSize*\myScaling] {\large $d_{\mathrm{e,d}}$} (ne5);
	\draw[-{Latex[length=1.5mm*\myScaling, width=1.1mm*\myScaling]}] (ne1) to (ne5);
	\draw[-{Latex[length=1.5mm*\myScaling, width=1.1mm*\myScaling]}] (ne2) to (ne3);
	\draw[-{Latex[length=1.5mm*\myScaling, width=1.1mm*\myScaling]}] (ne2) to (ne4);
	\draw[-{Latex[length=1.5mm*\myScaling, width=1.1mm*\myScaling]}] (ne2) to (ne5);
	\draw[-{Latex[length=1.5mm*\myScaling, width=1.1mm*\myScaling]}] (ne4) to (ne3);
	\draw[-{Latex[length=1.5mm*\myScaling, width=1.1mm*\myScaling]}] (ne4) to (ne5);

	\draw[-{Latex[length=1.5mm*\myScaling, width=1.1mm*\myScaling]}] (nt1) to node[auto,yshift=-0.1*\unitSize*\myScaling] {\large $\overbrace{T_{\mathrm{e,}1},\,V_{\mathrm{e,}1}}$} (nt2);
	\draw[-] (nt2) to (consumer);
	\draw[-{Latex[length=1.5mm*\myScaling, width=1.1mm*\myScaling]}] (nt3) to node[auto,yshift=0.1*\unitSize*\myScaling] {\large $\underbrace{T_{\mathrm{e,}3},\,V_{\mathrm{e,}3}}$} (nt4);
	\draw[-] (nt4) to (producer);
	\draw[-{Latex[length=1.5mm*\myScaling, width=1.1mm*\myScaling]}] (producer) to node[auto,xshift=0.5*\unitSize*\myScaling] {\large $T_{\mathrm{e,}4}$} (nt1);
	\draw[-{Latex[length=1.5mm*\myScaling, width=1.1mm*\myScaling]}] (consumer) to node[auto,xshift=-0.5*\unitSize*\myScaling] {\large $T_{\mathrm{e,}2}$} (nt3);
	\draw[fill=green,double distance=0.25mm*\myScaling,-{Latex[length=1.5mm*\myScaling, width=1.1mm*\myScaling]}] (power to heat) to node[auto,yshift=-0.1*\unitSize*\myScaling,xshift=-0.2*\unitSize*\myScaling] {\large $\dot{Q}_{\mathrm{pr,}1}$} (producer);
	\draw[double distance=0.25mm*\myScaling,-{Latex[length=1.5mm*\myScaling, width=1.1mm*\myScaling]}] (heat resource) to (power to heat);
	\draw[double distance=0.25mm*\myScaling,-{Latex[length=1.5mm*\myScaling, width=1.1mm*\myScaling]}] (consumer) to node[auto,yshift=0.4*\unitSize*\myScaling,xshift=-0.2*\unitSize*\myScaling] {\large $\dot{Q}_{\mathrm{d,}2}$} (heat sink);
	
	\draw[double distance=0.25mm*\myScaling,-{Latex[length=1.5mm*\myScaling, width=1.1mm*\myScaling]}] (hot pipe loss 1) to node[auto,xshift=-0.1*\unitSize*\myScaling,yshift=0.1*\unitSize*\myScaling] {\large $\dot{Q}_{\kappa,1}$} (hot pipe loss 2);
	\draw[double distance=0.25mm*\myScaling,-{Latex[length=1.5mm*\myScaling, width=1.1mm*\myScaling]}] (cold pipe loss 1) to node[auto,xshift=-0.65*\unitSize*\myScaling,yshift=-0.1*\unitSize*\myScaling] {\large $\dot{Q}_{\kappa,3}$} (cold pipe loss 2);

	\draw[-{Latex[length=1.5mm*\myScaling, width=1.1mm*\myScaling]}] (hot storage) to (nt1);
	\draw[-{Latex[length=1.5mm*\myScaling, width=1.1mm*\myScaling]}] (cold storage) to (nt4);

\end{tikzpicture}}
	\caption{Schematic of a minimal realization of an ETMG.}
	\label{fig:ETMG_figure}
\end{center}
\end{figure}

\textit{Notation:}
The symbols $\mathbb{R}$ and $\mathbb{N}$ denote the set of real numbers and nonnegative integers, respectively.
For a set $\mathcal V,$ $|\mathcal V|$ denotes its cardinality.
$1_{n}$ ($0_n$) denotes a vector containing $n$ ones (zeros).
The element-wise absolute value of a vector is $|x|$.
Let $X\in\mathbb{K}^{n\times m}$ and $0_{n\times m}$ be $n\times m$ matrices with all entries in $\mathbb{K}$ or equal to zero, respectively.
$I_n$ denotes the $n\times n$ identity matrix.
The notation $(\cdot)_i$ (or $(\cdot)_{i,j}$) refers to the element of the associated vector or matrix in the $i$-th row (and $j$-th column).
A matrix with the elements of a vector $x$ on its diagonal and zeros else is denoted by $\mathrm{diag}(x)$.
For the vectors $x\in\mathbb{K}^n,\,c\in\mathbb{K}^n$ we define $\norm{x}_{c}^2=x^{\mathrm{T}}\mathrm{diag}(c)x$.
A time dependence for a variable $x$ is denoted via $x(t)$ or $x(k)$ in continuous or discrete time, respectively. 

\section{Mathematical Modeling of Electro-Thermal Microgrids}\label{sec:ModelETMG}
We consider a generic ETMG and represent it by a multi-layer network, where each layer is associated with a physical domain, i.e., a thermal layer and an electrical layer.
Each layer consists of the respective grid and the units connected to it, i.e., the electrical and thermal storage units, producers and consumers.
The power-to-heat (P2H) connection of the two layers is implemented by heat pumps that couple nodes from the electrical layer with edges from the thermal layer.
Typically, the dynamics of the electrical layer and the heat pumps are fast compared to the thermal layer,
see \citep{hans_operation_2021,machado_modeling_2022}.
Hence, for the purpose of deriving an operation controller, we assume that the electrical grid as well as the units connected to it, with the exception of the state of charge of the electrical storage system (ESS), can be modeled by their steady-state equations.

In the following subsections we describe the structure of a generic ETMG by introducing the thermal and the electrical layer as well as the corresponding models for the grids, their units and the
heat pumps.
An illustrative, minimal realization of an ETMG is shown in Fig. \ref{fig:ETMG_figure}.

\subsection{Thermal layer}
We model the DHG of the ETMG by following the approaches from \cite{krug_nonlinear_2021} and \cite{machado_modeling_2022} via a connected graph ${\mathcal{G}_{\mathrm{t}}=(\mathcal{N}_{\mathrm{t}},\mathcal{E}_{\mathrm{t}})}$ with set of nodes $\mathcal{N}_{\mathrm{t}}\subset\mathbb{N}$ and set of edges $\mathcal{E}_{\mathrm{t}}\subset\mathbb{N}$.
The set $\mathcal{N}_{\mathrm{t}}$ is composed of ${n_{\mathrm{t,s}}\in\mathbb{N}}$ nodes representing thermal storage units as well as $n_{\mathrm{t,c}}\in\mathbb{N}$ nodes representing pipe crossings.
These nodes are interconnected via $|\mathcal{E}_{\mathrm{t}}|$ edges representing the $e_{\mathrm{t,t}}\in\mathbb{N}$ simple pipes and $e_{\mathrm{t,hp}}+e_{\mathrm{t,d}}$ heat exchanger, with the latter corresponding either to one of the $e_{\mathrm{t,hp}}\in\mathbb{N}$ heat pumps or one of the $e_{\mathrm{t,d}}\in\mathbb{N}$ thermal consumers.

We assign an arbitrary orientation to all edges and assume that it matches the corresponding flow direction of the water.
Later on, we show that this assumption is feasible for the derived hydraulic state dynamics.
The node-edge incidence matrix $F_{\mathrm{t}}\in\mathbb{R}^{|\mathcal{N}_{\mathrm{t}}|\times |\mathcal{E}_{\mathrm{t}}|}$ of $\mathcal{G}_{\mathrm{t}}$ is defined elementwise as
\begin{equation}\label{eq:F_t}
(F_{\mathrm{t}})_{i,j}=
	\begin{cases}
	1, & i\in \mathcal{N}_{\mathrm{t}}\text{ is sink of }j\in\mathcal{E}_{\mathrm{t}}, \\
	-1, & i\in \mathcal{N}_{\mathrm{t}}\text{ is source of }j\in\mathcal{E}_{\mathrm{t}}, \\
	0, & \text{otherwise.}
	\end{cases}
\end{equation}
By using this orientation and following \cite{machado_modeling_2022}, we define the sets $\mathcal{O}_{l}$ and $\mathcal{I}_{l}$ that collect all outcoming or incoming edges of node $l\in\mathcal{N}_{\mathrm{t}}$, i.e.,
\begin{equation*}
\begin{split}
\mathcal{O}_{l}&=\{i\in\mathcal{E}_{\mathrm{t}}\,|\,l\text{ is source of }i\in\mathcal{E}_{\mathrm{t}}\}, \\
\mathcal{I}_{l}&=\{i\in\mathcal{E}_{\mathrm{t}}\,|\,l\text{ is sink of }i\in\mathcal{E}_{\mathrm{t}}\}.
\end{split}
\end{equation*}
In what follows, $T_{\mathrm{e,}i}(t)\in\mathbb{R}$ and $x_{\mathrm{L},i}\in\begin{bmatrix} \underline{L},\overline{L} \end{bmatrix}\subset\mathbb{R}$ denote the average temperature and the longitudinal coordinate for all edges $i\in\mathcal{E}_{\mathrm{t}}$.
We associate a uniform temperature $T_{\mathrm{n,}l}(t)\in\mathbb{R}$ with each node $l\in\mathcal{N}_{\mathrm{t}}$.
Following common practice, see \cite{krug_nonlinear_2021,machado_modeling_2022}, the inflow temperature of all edges $i\in\mathcal{O}_l$ is set to the temperature of its source node, i.e., 
\begin{equation}\label{eq:Thermal_Cons_A}
	T_{\mathrm{e,}i}(\underline{L},t)=T_{\mathrm{n,}l}(t)\quad\forall i\in\mathcal{O}_l.
\end{equation}
Furthermore, we assume that the outflow temperature of each edge $i\in\mathcal{E}_{\mathrm{t}}$ is equal to its average temperature, i.e.,
\begin{equation}\label{eq:Thermal_Cons_B}
	T_{\mathrm{e,}i}(\overline{L},t)=T_{\mathrm{e,}i}(t).
\end{equation}
By (\ref{eq:Thermal_Cons_A}) and (\ref{eq:Thermal_Cons_B}), the spatial temperature profiles are omitted and a coherent temperature profile between all nodes and edges is ensured, see also e.g. \cite{krug_nonlinear_2021}.
\subsubsection{Thermal edges:}
The thermal edges, i.e., the simple pipes and heat exchanger are modeled as pipes transporting water as thermal energy carrier that exchanges heat flow with its environment, due to thermal losses, heat injection or extraction.
To describe the dynamics of the edges, we derive a physical model of water flow in pipes.
We assume a mainly one-dimensional flow of incompressible water throughout a DHG, which is not affected by gravitational forces.
This allows to model the flow of water in a pipe via the 1d incompressible Euler equations \citep{krug_nonlinear_2021}.
Additionally, the pumps of the network are considered non-controllable and to work in steady state at nominal operating conditions.
By applying the assumptions above to the 1d incompressible Euler equations with respect to the conservation of mass, we obtain \citep{krug_nonlinear_2021}
\begin{equation} \label{eq:1D-Euler_simpl_A}
\frac{\partial q_i(x_{\mathrm{L},i},t)}{\partial x_{\mathrm{L},i}} = 0, \;  \frac{\partial q_i(x_{\mathrm{L},i},t)}{\partial t}=0, \quad \forall i \in \mathcal{E}_{\mathrm{t}},
\end{equation}
for the volumetric flow $q_i\in\mathbb{R}$ of water.
Furthermore, it follows from the balance of momentum that
\begin{equation} \label{eq:1D-Euler_simpl_B}
\frac{\Delta p_i}{\overline{L}-\underline{L}} =-K_{\mathrm{f},i}\rho|q_i|q_i+\Delta p_{\mathrm{pump},i}, \, \forall i \in \mathcal{E}_{\mathrm{t}},
\end{equation}
holds for the pressure difference $\Delta p_i\in\mathbb{R}$ over edge $i$.
Therein, $K_{\mathrm{f},i}\in\mathbb{R}$ is the frictional coefficient, $\rho\in\mathbb{R}$ the density of water and $\Delta p_{\mathrm{pump},i}\in\mathbb{R}$ the delivered pressure difference, which is $\Delta p_{\mathrm{pump},i}>0$, if edge $i$ contains a pump and $\Delta p_{\mathrm{pump},i}=0$, otherwise.
From (\ref{eq:1D-Euler_simpl_A}) and (\ref{eq:1D-Euler_simpl_B}), a steady state operation of the hydraulics and flow dynamics follows.
Therefore, $q_i$ can be determined a priori.
This implies that the orientation of each edge $i\in\mathcal{E}_{\mathrm{t}}$ can be aligned according to the direction of the water flow $q_i$.

We regard the internal energy of water as the main energy carrier and neglect other energy forms. 
Furthermore, as in \cite{machado_modeling_2022}, we assume a linear dependency between the internal energy and the temperature of water.
The variables ${\dot{Q}_{\mathrm{pr,}i}(t)\in\mathbb{R}}$ and ${\dot{Q}_{\mathrm{d,}i}(t)\in\mathbb{R}}$ denote the induced and extracted heat flow at edge $i\in\mathcal{E}_{\mathrm{t}}$, respectively.
The heat loss at edge $i\in\mathcal{E}_{\mathrm{t}}$ is defined by
\begin{equation}
	\dot{Q}_{\kappa,i}(t)=\kappa_i\Big(T_{\mathrm{e,}i}(t)-T^{\mathrm{amb}}\Big),
\end{equation}
where $T_{\mathrm{e,}i}(t)$ is the temperature of the water flowing through the edge $i\in\mathcal{E}_{\mathrm{t}}$ and the parameters $T^{\mathrm{amb}}\in\mathbb{R}$ and $\kappa_i\in\mathbb{R}$ denote the ambient temperature and the heat loss coefficient  \citep{krug_nonlinear_2021}.
Thus, we model the temperature dynamics of edge $i\in\mathcal{O}_l,\,l\in\mathcal{N}_{\mathrm{t}}$ by 
\begin{align}
\rho cV_{\mathrm{e,}i}\frac{dT_{\mathrm{e,}i}(t)}{dt}=-\rho cq_{i}\big(T_{\mathrm{e,}i}(t)-&T_{\mathrm{n,}l}(t)\big)-\dot{Q}_{\kappa,i}(t)\nonumber\\
&+\dot{Q}_{\mathrm{pr,}i}(t)+\dot{Q}_{\mathrm{d,}i}(t). \label{eq:1D-Euler_simpl_C}
\end{align}
Therein, we assume constant values for the specific heat capacity $c\in\mathbb{R}$ and the volume of water $V_{\mathrm{e,}i}\in\mathbb{R}$ \citep{krug_nonlinear_2021}. 

\subsubsection{Thermal nodes:}
Each node of the thermal layer is associated with either a thermal storage unit or a pipe crossing.
Let $\mathcal{N}_{\mathrm{t,s}} \subset \mathcal{N}_{\mathrm{t}}$ denote the set of all thermal storage units.
For their temperature dynamics, we employ the temperature model from \cite{machado_modeling_2022}, i.e.,
\begin{equation}\label{eq:NodeTemp}
\rho cV_{\mathrm{n,}l}\frac{dT_{\mathrm{n,}l}(t)}{dt}=\rho c\Big(\sum_{i\in \mathcal{I}_l}q_{i}T_{\mathrm{e,}i}(t)-\sum_{i\in \mathcal{O}_l}q_{i}T_{\mathrm{n,}l}(t)\Big),
\end{equation}
for all $l\in\mathcal{N}_{\mathrm{t,s}}$.
Put another way, (\ref{eq:NodeTemp}) models the nodal temperature change according to the in- and outflow temperatures with respect to their volumetric flow and the node volume $V_{\mathrm{n},l}$.

Last but not least, the set $\mathcal{N}_{\mathrm{t,c}}\subset\mathcal{N}_{\mathrm{t}}$ denotes the set of all pipe crossings.
We assume that their volume is negligible compared to those of the pipes and the thermal storage units, i.e., $V_{\mathrm{n,}l}=0\,\forall\,l\in\mathcal{N}_{\mathrm{t,c}}$.

\subsubsection{Discrete-time state model for thermal layer:}
In what follows, we derive a model of the thermal layer of the form
\begin{equation} \label{eq:ThermalStateSpace}
	x_{\mathrm{t}}(k+1) =  A_{\mathrm{t}}x_{\mathrm{t}}(k) + B_{\mathrm{t}}u_{\mathrm{t}}(k) + E_{\mathrm{t}}d_{\mathrm{t}}(k).
\end{equation}
To this end, we collect the temperatures of all edges and thermal storage units in $T_{\mathrm{e}}(t)\in\mathbb{R}^{n_{\mathrm{t,e}}}$ and $T_{\mathrm{n,s}}(t)\in\mathbb{R}^{n_{\mathrm{t,s}}}$, respectively, to form the state vector $$x_{\mathrm{t}}(t)=[T_{\mathrm{e}}^{\mathrm{T}}(t)\;T_{\mathrm{n,s}}^{\mathrm{T}}(t)]^{\mathrm{T}}.$$
The sets of the $e_{\mathrm{t,hp}}$ heat pumps and $e_{\mathrm{t,d}}$ thermal consumers are denoted $\mathcal{E}_{\mathrm{t,hp}}\subset \mathcal{E}_{\mathrm{t}}$ and $\mathcal{E}_{\mathrm{t,d}}\subset \mathcal{E}_{\mathrm{t}}$, respectively.
This allows to define the control signal vector containing the heat flows produced by the heat pumps, i.e.,
\begin{equation*}
	u_{\mathrm{t}}(t)=\begin{bmatrix}\dot{Q}_{\mathrm{pr},e_{\mathrm{pr,}1}}(t) \cdots \dot{Q}_{\mathrm{pr},e_{\mathrm{pr,}n}}(t)\end{bmatrix}^{\mathrm{T}}\in\mathbb{R}^{e_{\mathrm{t,hp}}},
\end{equation*} 
where $\{e_{\mathrm{pr,}1},\ldots,e_{\mathrm{pr,}n}\}=\mathcal{E}_{\mathrm{t,hp}}$.
Moreover, the disturbance vector containing the consumed heat flows is
\begin{equation*}
	d_{\mathrm{t}}(t)=\begin{bmatrix}\dot{Q}_{\mathrm{d},e_{\mathrm{d,}1}}(t) \cdots \dot{Q}_{\mathrm{d},e_{\mathrm{d,}n}}(t),T^{\mathrm{amb}}\end{bmatrix}^{\mathrm{T}}\in\mathbb{R}^{e_{\mathrm{t,d}}+1},
\end{equation*}
where $\{e_{\mathrm{d,}1},\ldots,e_{\mathrm{d,}n}\}=\mathcal{E}_{\mathrm{t,d}}$.

To form (\ref{eq:ThermalStateSpace}), we divide the incidence matrix $F_{\mathrm{t}}$ into a positive and a negative part, see \cite{machado_modeling_2022},
i.e., $F_{\mathrm{t}}^{+}=\frac{1}{2}(F_{\mathrm{t}}+|F_{\mathrm{t}}|)$, $F_{\mathrm{t}}^{-}=\frac{1}{2}(|F_{\mathrm{t}}|-F_{\mathrm{t}})$.

Since the temperatures of the nodes $l\in\mathcal{N}_{\mathrm{t,c}}$ are not part of $x_{\mathrm{t}}(t)$, we eliminate all rows $(F_{\mathrm{t}}^{+})_l,\,(F_{\mathrm{t}}^{-})_l,\,{l\in\mathcal{N}_{\mathrm{t,c}}}$ from $F_{\mathrm{t}}^{+}$ and $F_{\mathrm{t}}^{-}$ to obtain $\tilde{F}_{\mathrm{t}}^{+}\in\mathbb{N}^{n_{\mathrm{t,s}}\times |\mathcal{E}_{\mathrm{t}}|}$ and ${\tilde{F}_{\mathrm{t}}^{-}\in\mathbb{N}^{n_{\mathrm{t,s}}\times |\mathcal{E}_{\mathrm{t}}|}}$.
For what follows, it is useful to define the matrix ${\tilde{A}_{\mathrm{t}}\in\mathbb{R}^{|\mathcal{E}_{\mathrm{t}}|\times |\mathcal{E}_{\mathrm{t}}|}}$ elementwise as 
\begin{equation*}
	(\tilde{A}_{\mathrm{t}})_{i,j}=
	\begin{cases}
	-q_i-\frac{\kappa_i}{\rho c},\quad\textrm{if }i=j,\\[10pt]
	\begin{aligned}
	\frac{q_i q_j}{\sum_{l\in\mathcal{O}_{l}}q_l},\quad\textrm{if } \exists \, l&\in\mathcal{N}_{\mathrm{t,c}}, \\\\[-20pt]
	&\text{s.t. }(F_{\mathrm{t}}^{+})_{l,j}=(F_{\mathrm{t}}^{-})_{l,i}=1,
	\end{aligned} \\
	0,\quad\text{otherwise.}
	\end{cases}
\end{equation*}
We define $J_{\mathrm{t}}=\rho c\,\mathrm{diag}(\begin{bmatrix}
V_{\mathrm{e}}^{\mathrm{T}}\;V_{\mathrm{n,s}}^{\mathrm{T}}
\end{bmatrix}^{\mathrm{T}})$ with $V_{\mathrm{e}}\in\mathbb{R}^{|\mathcal{E}_{\mathrm{t}}|}$ and $V_{\mathrm{n,s}}\in\mathbb{R}^{n_{\mathrm{t,s}}}$ as thermal inertia matrix collecting the volumes of all edges and all thermal storage units, respectively.
Finally, we are able to define the system matrix
\begin{equation*}
    A_{\mathrm{t}}'=\rho cJ_{\mathrm{t}}^{-1}\begin{bmatrix}
	\tilde{A}_{\mathrm{t}} & \mathrm{diag}(|q_{\mathrm{t}}|)(\tilde{F}_{\mathrm{t}}^{-})^{\mathrm{T}} \\
	\tilde{F}_{\mathrm{t}}^{+}\mathrm{diag}(|q_{\mathrm{t}}|) & -\mathrm{diag}(\tilde{F}_{\mathrm{t}}^{+}|q_{\mathrm{t}}|)
	\end{bmatrix},
\end{equation*}
where $q_{\mathrm{t}}\in\mathbb{R}^{|\mathcal{E}_{\mathrm{t}}|}$ comprises the constant volumetric flows.

The control matrix $B_{\mathrm{t}}'\in\mathbb{R}^{(|\mathcal{E}_{\mathrm{t}}|+n_{\mathrm{t,s}})\times e_{\mathrm{t,hp}}}$ and the disturbance matrix $E_{\mathrm{t}}'\in\mathbb{R}^{(|\mathcal{E}_{\mathrm{t}}|+n_{\mathrm{t,s}})\times(e_{\mathrm{t,d}}+1)}$ are defined as
\begin{equation*}
    B_{\mathrm{t}}'=J_{\mathrm{t}}^{-1}\begin{bmatrix} \tilde{B}_{\mathrm{t}}^{\mathrm{T}} & 0_{n_{\mathrm{t,s}}\times e_{\mathrm{t,hp}}}^{\mathrm{T}} \end{bmatrix}^{\mathrm{T}},\,
    E_{\mathrm{t}}'=J_{\mathrm{t}}^{-1}\begin{bmatrix} \begin{matrix} \tilde{E}_{\mathrm{t}} & \kappa \end{matrix} \\ 0_{n_{\mathrm{t,s}}\times (e_{\mathrm{t,d}}+1)} \end{bmatrix},
\end{equation*}
where $\tilde{B}_{\mathrm{t}}\in\mathbb{N}^{|\mathcal{E}_{\mathrm{t}}|\times|\mathcal{E}_{\mathrm{t,hp}}|}$ ($\tilde{E}_{\mathrm{t}}\in\mathbb{N}^{|\mathcal{E}_{\mathrm{t}}|\times|\mathcal{E}_{\mathrm{t,d}}|}$) with $\big(\tilde{B}_{\mathrm{t}}\big)_{i,j}=1$, if there exists an $i\in\mathcal{E}_{\mathrm{t}}$ and $j\in\mathcal{E}_{\mathrm{t,hp}}$ ($j\in\mathcal{E}_{\mathrm{t,d}}$), such that $i=j$, and $\kappa\in\mathbb{R}^{|\mathcal{E}_{\mathrm{t}}|}$ collecting all heat loss coefficients.
The matrices $A_{\mathrm{t}}'$, $B_{\mathrm{t}}'$ and $E_{\mathrm{t}}'$ are defined for the system dynamics in continuous-time.
For our MPC scheme, we derive a discrete-time model of the form (\ref{eq:ThermalStateSpace}) by calculating matrix exponentials for a fixed sample time $\Delta t$, which results in $A_{\mathrm{t}}$, $B_{\mathrm{t}}$ and $E_{\mathrm{t}}$ in (\ref{eq:ThermalStateSpace}).

\subsection{Electrical layer}
Following \cite{schiffer2016survey}, we represent the topology of the electrical AC power network as a weighted, undirected, connected graph $\mathcal{G}_{\mathrm{e}}=(\mathcal{N}_{\mathrm{e}},\mathcal{E}_{\mathrm{e}},\mathcal{W}_{\mathrm{e}})$ with set of nodes $\mathcal{N}_{\mathrm{e}}\subset\mathbb{N}$, set of edges $\mathcal{E}_{\mathrm{e}}\subset\mathbb{N}$ and set of edge weights $\mathcal{W}_{\mathrm{e}}\subset\mathbb{R}.$
One of the $|\mathcal{N}_{\mathrm{e}}|$ nodes represents the point of common coupling (PCC) 
that connects the electrical layer of the ETMG with the main transmission grid. All other nodes represent electrical units.
These are $n_{\mathrm{e,r}}\in\mathbb{N}$ RES, $n_{\mathrm{t,hp}}\in\mathbb{N}$ heat pumps, $n_{\mathrm{e,d}}\in\mathbb{N}$ loads, and $n_{\mathrm{e,s}}\in\mathbb{N}$ ESS.
The vectors $d_{\mathrm{e,r}}\in\mathbb{R}^{n_{\mathrm{e,r}}}$, $d_{\mathrm{e,d}}\in\mathbb{R}^{n_{\mathrm{e,d}}}$, $u_{\mathrm{e,hp}}\in\mathbb{R}^{n_{\mathrm{e,hp}}}$, $u_{\mathrm{e,s}}\in\mathbb{R}^{n_{\mathrm{e,s}}}$, and the scalar $u_{\mathrm{e,t}}\in\mathbb{R}$ denote the electrical power provided or consumed by RES, loads, heat pumps, ESS and the main grid, respectively.
We number the nodes in ascending order so that the first node is associated to the PCC, followed by the connections to the ESS, the P2H units, the RES and the loads.
This allows to define 
\begin{equation} \label{eq:Def_p_e,n}
	p_{\mathrm{e,n}}=\begin{bmatrix}u_{\mathrm{e,t}} & u_{\mathrm{e,s}}^{\mathrm{T}} & u_{\mathrm{e,hp}}^{\mathrm{T}} & d_{\mathrm{e,r}}^{\mathrm{T}} & d_{\mathrm{e,d}}^{\mathrm{T}} \end{bmatrix}^{\mathrm{T}}\in\mathbb{R}^{|\mathcal{N}_{\mathrm{e}}|},
\end{equation}
which collects all power flows that are exchanged between units and nodes.
The nodes are interconnected via $|\mathcal{E}_{\mathrm{e}}|$ edges representing power lines.
We again choose an arbitrary orientation for each edge $i\in\mathcal{E}_{\mathrm{e}}$ (see, e.g., Fig. \ref{fig:ETMG_figure}) and construct the incidence matrix $F_{\mathrm{e}}\in\mathbb{N}^{|\mathcal{N}_{\mathrm{e}}|\times |\mathcal{E}_{\mathrm{e}}|}$ of the electrical network analogously to (\ref{eq:F_t}).

\subsubsection{Electrical edges:}
We assume the power system is operated under symmetric conditions, see \cite{schiffer2016survey}.
To model the power transfer over all edges $i\in\mathcal{E}_{\mathrm{e}}$, we follow \cite{hans_operation_2021} and assume short power lines that are purely inductive and have no shunt admittances.
Thus, the absolute value of each series susceptance, i.e., the edge weight $w_i=b_{l,m}=b_{l,m}\in\mathcal W$, is associated with the power line $i\in\mathcal{E}_{\mathrm{e}}$ connecting nodes $m$ and $l$.
Additionally, constant voltage amplitudes $\hat{v}_{m}\in\mathbb{R}$ are assumed for all $m\in\mathcal{N}_{\mathrm{e}}$, which allow us to merge the susceptances and voltage amplitudes to the parameter $a_i=b_{m,l}\hat{v}_{m}\hat{v}_{l}$.
The vector $a\in\mathbb{R}^{|\mathcal{E}_{\mathrm{e}}|}$ collects all parameters $a_i$.
The difference of the phase angles $\Theta_{m}\in\mathbb{R},\,m\in\mathcal{N}_{\mathrm{e}},$ between the units in the electrical grid are assumed to be small, such that $\sin(\Theta_{m}(t)-\Theta_{l}(t))\approx \Theta_{m}(t)-\Theta_{l}(t)$ and $\cos(\Theta_{m}(t)-\Theta_{l}(t))\approx 1,\, \forall(m,l)\in\mathcal{N}_{\mathrm{e}}$.
 
The foregoing assumptions allow us to use an approximate DC power flow model to describe the active power flows in the electrical layer.
As discussed in \cite{purchala_usefulness_2005}, this approximation typically does not cause unacceptable model mismatch, even for small reactance to resistance ratios. 
To derive the electrical power transmitted over the electrical lines $p_{\mathrm{e,e}}\in\mathbb{R}^{|\mathcal{E}_{\mathrm{e}}|}$ directly from $p_{\mathrm{e,n}}$, we employ
\begin{equation*}
	\begin{split}
	T&=\begin{bmatrix}
		I_{(n_{\mathrm{e,n}}-1)} & -1_{(n_{\mathrm{e,n}}-1)} \\
		0_{(n_{\mathrm{e,n}}-1)}^{\mathrm{T}} & 1
	\end{bmatrix}, \quad
	\tilde{T}=\begin{bmatrix} 
		I_{(n_{\mathrm{e,n}}-1)} & 0_{(n_{\mathrm{e,n}}-1)}
	 \end{bmatrix},\\
	\mathcal{L}&=F_{\mathrm{e}}\mathrm{diag}(a)F_{\mathrm{e}}^{\mathrm{T}}, \quad
	\tilde{\mathcal{L}}=\tilde{T}\mathcal{L}\tilde{T}^{\mathrm{T}}, \\
	\tilde{F}_{\mathrm{e}}&=\mathrm{diag}(a)F_{\mathrm{e}}^{\mathrm{T}}T^{-1}\tilde{T}^{\mathrm{T}}\tilde{\mathcal{L}}^{-1}\tilde{T},
\end{split} 
\end{equation*}
from \cite{hans_operation_2021}.
Then, $\tilde{F}_{\mathrm{e}}$ allows us to model $p_{\mathrm{e,e}}(k)$ via
\begin{equation}\label{eq:BalanceElectricalLayerI}
	p_{\mathrm{e,e}}(k)=\tilde{F}_{\mathrm{e}} \, p_{\mathrm{e,n}}(k),
\end{equation}
which describes the connection between the provided or extracted power at the nodes and the power flow over the power lines.
Moreover, a global power balance must hold, i.e., 
\begin{equation}\label{eq:BalanceElectricalLayerII}
	0=p_{\mathrm{e,n}}(k) \, 1_{|\mathcal{N}_{\mathrm{e}}|}.
\end{equation}

\subsubsection{Electrical nodes:}
Each node of the electrical layer is either associated with a unit or represents the PCC with the main grid.
The power exchange at the PCC, i.e., $u_{\mathrm{e,t}}(k)$, is regarded as control signal.
The load demand $d_{\mathrm{e,d}}(k)$ and RES infeed $d_{\mathrm{e,r}}(k)$, are assumed to be known for all $k\in\mathbb{N}$.

We model the states of charge $x_{\mathrm{e}}(t)\in\mathbb{R}^{n_{\mathrm{e,s}}}$ of the ESS by integrating its power $u_{\mathrm{e,s}}(t)\in\mathbb{R}^{n_{\mathrm{e,s}}}$ over time.
By using the sample time $\Delta t$ and applying an exact discretization method, we obtain the discrete-time state model
\begin{equation} \label{eq:ElectricalStateSpace}
	x_{\mathrm{e}}(k+1)=x_{\mathrm{e}}(k)+B_{\mathrm{e}}u_{\mathrm{e,s}}(k),
\end{equation}
with $B_{\mathrm{e}}=I_{n_{\mathrm{e,s}}}\Delta t$.

\subsection{Power-to-heat interconnection}
The P2H interconnection is incorporated by heat pumps connected to the nodes $\mathcal{N}_{\mathrm{e,hp}}\subset\mathcal{N}_{\mathrm{e}}$.
At each node $l\in\mathcal{N}_{\mathrm{e,hp}}$, a heat pump transforms the extracted electrical power $(u_{\mathrm{e,hp}}(k))_l$ into heat flow $(u_{\mathrm{t}}(k))_i$, which is then induced at the corresponding thermal edge $i\in\mathcal{E}_{\mathrm{t,hp}}$.
By assuming steady state operation, we model heat pumps via \citep{carli_robust_2022}
\begin{equation} \label{eq:PowerToHeatUnit}
	u_{\mathrm{t}}(k)=-\mathrm{diag}(\alpha)\,u_{\mathrm{e,hp}}(k),
\end{equation}
with $\alpha\in\mathbb{R}^{n_{\mathrm{t,hp}}}$ collecting the coefficients of performance.

\subsection{Overall ETMG model}
In order to obtain a compact discrete-time model of the ETMG, we define the state, disturbance and control vectors as
\begin{equation*}
	\begin{split} 
x(k)&=\begin{bmatrix} x_{\mathrm{e}}^{\mathrm{T}}(k) & x_{\mathrm{t}}^{\mathrm{T}}(k) \end{bmatrix}^{\mathrm{T}}, d_{\mathrm{e}}(k)=\begin{bmatrix}d_{\mathrm{e,r}}^{\mathrm{T}}(k) & d_{\mathrm{e,d}}^{\mathrm{T}}(k)\end{bmatrix},\\
u_{\mathrm{e}}(k)&=\begin{bmatrix} u_{\mathrm{e,t}}(k) & u_{\mathrm{e,s}}^{\mathrm{T}}(k) & u_{\mathrm{e,hp}}^{\mathrm{T}}(k)\end{bmatrix}^{\mathrm{T}},
\end{split} 
\end{equation*}
and combine \eqref{eq:ThermalStateSpace} and \eqref{eq:BalanceElectricalLayerI} - \eqref{eq:PowerToHeatUnit} to
\begin{equation}
	\begin{cases}
		x(k+1)&=A x(k)+B u_{\mathrm{e}}(k)+E d_{\mathrm{t}}(k), \\
		0_{|\mathcal{N}_{\mathrm{e}}|} &= p_{\mathrm{e,e}}(k)-\tilde{F}_{\mathrm{e}}\begin{bmatrix}
			u_{\mathrm{e}}^{\mathrm{T}}(k) & d_{\mathrm{e}}^{\mathrm{T}}(k)
		\end{bmatrix}^{\mathrm{T}}, \\
	0&=\begin{bmatrix}u_{\mathrm{e}}^{\mathrm{T}}(k) & d_{\mathrm{e}}^{\mathrm{T}}(k)\end{bmatrix}^{\mathrm{T}}\cdot 1_{|\mathcal{N}_{\mathrm{e}}|}
	\end{cases}	
	\label{etmg_model}
\end{equation}
where
\begin{equation*}
	\begin{split}
	A&=\begin{bmatrix}
		I_{n_{\mathrm{e,s}}} & 0_{n_{\mathrm{e,s}}\times(|\mathcal{E}_{\mathrm{t}}|+n_{\mathrm{t,s}})} \\
		0_{(|\mathcal{E}_{\mathrm{t}}|+n_{\mathrm{t,s}})\times n_{\mathrm{e,s}}} & A_{\mathrm{t}}
	\end{bmatrix}, \\
	B&=\begin{bmatrix} 
		0_{n_{\mathrm{e,s}}} & B_{\mathrm{e}} & 0_{n_{\mathrm{e,s}}\times e_{\mathrm{t,hp}}} \\
		0_{(|\mathcal{E}_{\mathrm{t}}|+n_{\mathrm{t,s}})} & 0_{(|\mathcal{E}_{\mathrm{t}}|+n_{\mathrm{t,s}})\times n_{\mathrm{e,s}}} & -B_{\mathrm{t}}\mathrm{diag}(\alpha) \end{bmatrix}, \\
		E&=\begin{bmatrix}
		0_{n_{\mathrm{e,s}}\times (e_{\mathrm{t,d}}+1)} \\
		E_{\mathrm{t}}
	 \end{bmatrix}.
\end{split} 
\end{equation*}

\section{Model Predictive Operation Control algorithm for ETMGs}\label{sec:OptimizationModel}
Our goal is to operate the ETMG employing MPC, such that an increased operating flexibility
 is achieved, while electrical and heat demand as well as operational constraints are met.
The core of an MPC algorithm is an optimization problem which is used to 
recurrently determine optimized control signals by minimizing an objective function subject to constraints.
While the model from Section~\ref{sec:ModelETMG} enters the OP via equality constraints, the objective function and further constraints defining admissible operating regions are introduced in Sections~\ref{subsec:ObjectiveFunktion} and \ref{subsec:Constraints}.
Then, the MPC problem is posed in Section \ref{subsec:OptimizationModel}.

 

\subsection{Objective function} \label{subsec:ObjectiveFunktion}
The control objective is given by a function that penalizes deviations from the optimal operation over the entire prediction horizon \citep{morari_model_1999}.
We therefore define different operating stage costs at time instant $k$, which are then combined into one objective function $J$.

The first part is motivated by a state-of-the-art supply temperature-based operation of a DHG, which is relevant for, e.g., temperature sensitive loads.
This requires the temperature of particular nodes or edges of $\mathcal{G}_{\mathrm{t}}$, which we refer to as manipulated temperatures, to be maintained close to a desired temperature.
Therefore, we derive a stage cost that penalizes deviations from the desired value.
For this, we define the desired temperature of node or edge $l$ as $x_{\mathrm{t},l}^{\mathrm{d}}\in\mathbb{R}$.
Furthermore, let $\mathcal{X}_{\mathrm{t}}\subset(\mathcal{N}_{\mathrm{t}}\cup\mathcal{E}_{\mathrm{t}})$ denote the set of all nodes or edges to be operated close to their corresponding $x_{\mathrm{t},l}^{\mathrm{d}}$.
This allows us to define the vector collecting all manipulated temperatures 
\begin{equation*}
	x_{\mathrm{t,mp}}=\begin{bmatrix} (x_{\mathrm{t}})_{n_{\mathrm{t,}1}} & \cdots & (x_{\mathrm{t}})_{n_{\mathrm{t,}n}} \end{bmatrix}^{\mathrm{T}}\in\mathbb{R}^{|\mathcal{X}_{\mathrm{t}}|},
\end{equation*}
where $\{n_{\mathrm{t,}1},\ldots,n_{\mathrm{t,}n}\}=\mathcal{X}_{\mathrm{t}}$.
Let all desired supply temperatures be collected in the vector $x_{\mathrm{t}}^{\mathrm{d}}\in\mathbb{R}^{|\mathcal{X}_{\mathrm{t}}|}$.
Then, we can formulate the cost
\begin{equation} \label{eq:tempStageCost}
	l_{\mathrm{t}} (k) = \norm{x_{\mathrm{t,mp}}(k)-x_{\mathrm{t}}^{\mathrm{d}}}_{c_{\mathrm{t}}}^2,
\end{equation}
with weight $c_{\mathrm{t}}\in\mathbb{R}^{|\mathcal{X}_{\mathrm{t}}|}$, to penalize deviations of $x_{\mathrm{t,mp}}(k)$ from $x_{\mathrm{t}}^{\mathrm{d}}$.

Second, we consider economically motivated costs aimed to minimize the power consumption of the controllable ETMG components.
The corresponding stage cost reads
\begin{equation} \label{eq:economicStageCost}
	l_{\text{ec,}i} (k) = \norm{u_{\mathrm{e},i}(k)}_{c_{\mathrm{e},i}}^2,
\end{equation}
where $c_{\mathrm{e},i}\in\mathbb{R}^{n_{\mathrm{e,}i}}$ is a weight that allows to define operational preferences of specific parts for $i\in\{\mathrm{t,\,s,\,hp}\}$ and $n_{\mathrm{e,hp}}=n_{\mathrm{t,hp}}$.


To account for power dependent varying efficiencies of heat pumps while $\alpha$ in \eqref{eq:PowerToHeatUnit} is constant, we assume a best efficiency operating point $u_{\mathrm{hp}}^{\mathrm{d}}\in\mathbb{R}^{n_{\mathrm{e,hp}}}$ \citep{arpagaus_hochtemperatur-warmepumpen_2019}.
Deviation of the operating power $u_{\mathrm{e,hp}}(k)$ from $u_{\mathrm{hp}}^{\mathrm{d}}$ results in increased conversion losses.
Additionally, a highly dynamic operation of a heat pump is considered detrimental in terms of its service life.
Therefore, an efficiency-based stage costs is introduced as
\begin{equation} \label{eq:effStageCost}
	\begin{aligned}
	l_{\mathrm{hp}}(k) =& \norm{u_{\mathrm{e,hp}}(k)-u_{\mathrm{hp}}^{\mathrm{d}}}^2_{c_{\mathrm{hp,I}}} \\
	+&\norm{u_{\mathrm{e,hp}}(k)-u_{\mathrm{e,hp}}(k-1)}^2_{c_{\mathrm{hp,II}}},
	\end{aligned}
\end{equation}
with weights $c_{\mathrm{hp,I}}\in\mathbb{R}^{n_{\mathrm{t,hp}}}$, $c_{\mathrm{hp,II}}\in\mathbb{R}^{n_{\mathrm{t,hp}}}$ and the previous control signal $u_{\mathrm{e,hp}}(k-1)$.

Combining (\ref{eq:tempStageCost}) to (\ref{eq:effStageCost}), we are now able to define the quadratic objective function 
\begin{equation}
	J=\sum_{k\in\mathbb{N}_{\mathrm{h}}} l_{\mathrm{t}} (k) + l_{\mathrm{ec,t}} (k) + l_{\mathrm{ec,s}} (k) + l_{\mathrm{ec,hp}} (k) + l_{\mathrm{hp}}(k),
	\label{J}
\end{equation}
with $\mathbb{N}_{\mathrm{h}}=\{k_{0},k_{0}+1,\ldots,k_{0}+N-1\}$ and initial time step $k_{0}\in\mathbb{N}$ based on the costs over prediction horizon $N$.

\subsection{Admissible operating regions} \label{subsec:Constraints}
Admissible operating regions for the dynamic states $x_{\mathrm{t}}(k)$ and $x_{\mathrm{e}}(k)$, the power flow over the power lines $p_{\mathrm{e,e}}(k)$ and the control signal $u_{\mathrm{e}}(k)$ are defined below.
For this $\underline{(\cdot)}$ and $\overline{(\cdot)}$ denote the lower and upper operating limits of the associated quantity in $\mathbb{R}$, respectively. 

The admissible set of flow temperatures $\mathbb{X}_{\mathrm{t}} \subset \mathbb{R}^{(|\mathcal{E}_{\mathrm{t}}|+n_{\mathrm{t,s}})}$ is defined as
\begin{equation*}
	\mathbb{X}_{\mathrm{t}}=\{\underline{x}_{\mathrm{t,}l}\leq (x_{\mathrm{t}})_l\leq \overline{x}_{\mathrm{t,}l},l=1,...,(|\mathcal{E}_{\mathrm{t}}|+n_{\mathrm{t,s}})\}.
\end{equation*}
It follows that the temperatures $x_{\mathrm{t}}(k)$ can flexibly adapt according to the heat demand and supply within $\mathbb{X}_{\mathrm{t}}$.

The capacities of the $l$-th ESS, $\overline{x}_{\mathrm{e,}l}$, determine its admissible operating regions $\mathbb{X}_{\mathrm{e}}\subset\mathbb{R}^{n_{\mathrm{e,s}}}$, i.e.,
\begin{equation*} \label{eq:SoCConstraints}
	\mathbb{X}_{\mathrm{e}}=\{0\leq(x_{\mathrm{e}})_l\leq \overline{x}_{\mathrm{e,}l},l=1,...,n_{\mathrm{e,s}}\}.
\end{equation*} 
Furthermore, we define $\mathbb{X}=\mathbb{X}_{\mathrm{t}} \times \mathbb{X}_{\mathrm{e}}\subset\mathbb{R}^{(|\mathcal{E}_{\mathrm{t}}|+n_{\mathrm{t,s}}+n_{\mathrm{e,s}})}$.

The electrical power flow over the lines and the control signals are constrained by
\begin{align*}
	\mathbb{P}_{\mathrm{e}}&=\{\underline{p}_{\mathrm{e,}l}\leq(p_{\mathrm{e,e}})_l\leq \overline{p}_{\mathrm{e,}l},l=1,...,n_{\mathrm{e,e}}\},\text{ and} \\
	\mathbb{U}_{\mathrm{e}}&=\{\underline{u}_{\mathrm{e,}i,l}\leq(u_{\mathrm{e,}i})_l\leq \overline{u}_{\mathrm{e,}i,l},l=1,...,n_{\mathrm{e,}i}\},i\in\{\mathrm{t,\,s,\,hp}\},
\end{align*}
to account for performance limitations and prevent overloads of power lines or controllable units.

\subsection{Model predictive operation controller} \label{subsec:OptimizationModel}
Besides the objective function and the constraints, the disturbance inputs need to be incorporated into the MPC algorithm.
Since we assume perfect forecasts, $d_{\mathrm{t}}(k)$ and $d_{\mathrm{e}}(k)$ are considered known for all $k\in\mathbb{N}_{\mathrm{h}}$. 
Although this assumption is violated in many real-world applications, dealing with uncertainties in power or heat demand and supply in a structured manner is out of scope of this paper.

Under this assumption, and after defining the objective function and the constraints, we can now formulate a convex MPC problem, using $U_{\mathrm{e}}=\begin{bmatrix}u_{\mathrm{e}}(k_0),\ldots,u_{\mathrm{e}}(k_0+N-1)
\end{bmatrix}$.\\

\begin{problem}[MPC for ETMG]\label{pr:OP} Find the optimal control sequence $U_{\mathrm{e}}^*=\begin{bmatrix}u_{\mathrm{e}}^*(k_0),\ldots,u_{\mathrm{e}}^*(k_0+N-1)
\end{bmatrix}$ that solves
	\begin{equation*}
    \begin{aligned}
        \min_{U_{\mathrm{e}}} \;&J \\
        \text{s.t. }&(\ref{etmg_model}), \, x(k)\in\mathbb{X}, \, p_{\mathrm{e,e}}(k)\in\mathbb{P}_{\mathrm{e}}, \, u_{\mathrm{e}}(k)\in\mathbb{U}_{\mathrm{e}}, 
    \end{aligned}
\end{equation*}
for given $k_{0}$, $x(k_{0})$, $u_{\mathrm{e,hp}}(k_{0}-1)$ and known $d_{\mathrm{t}}(k)$ and $d_{\mathrm{e}}(k)$ for all $k\in\mathbb{N}_{\mathrm{h}}$.
\end{problem}

For the MPC algorithm, Problem \ref{pr:OP} is implemented in a receding horizon fashion, i.e., only $u_{\mathrm{e}}^*(k_{0})$ is applied to the ETMG at time $t=k_0.$
At the next execution, Problem \ref{pr:OP} is solved again with updated initial values for state and uncertain input vectors. Since Problem \ref{pr:OP} is quadratic and convex, existing numerical methods can be used to compute a solution efficiently.

\section{Case study}\label{sec:CaseStudy}
The performance of the predictive operation controller derived in Section~\ref{subsec:OptimizationModel} is now illustrated for the ETMG shown in Fig.~\ref{fig:ETMG_figure}.
The goal is to analyze the extent to which the proposed controller can reduce economic and efficiency-based operating costs, while increasing the flexibility in operation.
For this purpose, the ETMG model \eqref{etmg_model} and the MPC algorithm from Section~\ref{subsec:OptimizationModel} using two different sets of cost weights, resulting in two scenarios, are investigated. 
We use the superscript (I or II) to associate the variables to each of the scenarios.
For the case study, MPC I serves as a benchmark while MPC II aims to investigate the achievable potential gains mentioned above.

The investigated ETMG comprises all units as introduced in Section \ref{sec:ModelETMG}.
Since we focus on an operation controller that exploits variable flow temperatures, we assume equal admittances $\tilde{a}$ for all lines, i.e., $a=\tilde{a}\,1_4$, to simplify the model of the electrical layer and discuss the setup of the thermal layer in greater detail.
The pipes of the thermal layer have the length $L=5\unit{km}$ with diameter $D=0.1\unit{m}$, which corresponds to typical distances between central heat generators and consumers in urban areas.
We define a nominal heat flow $\dot{Q}_{\mathrm{N}}=3\unit{MW_{th}}$, a supply temperature $T_{\mathrm{N}}=90^{\circ}\unit{C}$ as well as a temperature difference between supply and return pipes of $\Delta T=30K$.
The heat loss coefficient for all pipes is set, such that $5\%$ of $\dot{Q}_{\mathrm{N}}$ dissipate, if the pipe is operated at $T_{\mathrm{N}}$. 
The remaining model parameters are $T^{\mathrm{amb}}=10\,\unit{^{\circ}C}$, $c = 4182\,\unit{\frac{J}{kg\,K}}$, $\rho = 987\,\unit{\frac{kg}{m^3}}$ and $V_{\mathrm{n,s}} = 100\,\unit{m^3}$.

\begin{figure}[!h]

\colorlet{colorDemandHeat}{vgDarkBlue}
\colorlet{colorStorage}{vgOrange}
\colorlet{colorRes}{vgGreen}
\colorlet{colorDemandPower}{vgRed}
\colorlet{colorGrid}{vgOrangeII}
\colorlet{colorP2H_el}{vgLightBlue}
\colorlet{colorP2H_th}{vgYellow}
\colorlet{colorQ_loss}{vgDeepBlue}

\pgfplotsset{limit/.style = {const plot, color=black, dash pattern=on 1pt off 3pt on 3pt off 3pt, thick}}

\newcommand{\PlotLimit}[1]{%
  \addplot[limit, forget plot] plot coordinates{
    (1,#1)
    (4*4*24,#1)
  };
}

\pgfplotsset{resultsPlot/.style={%
	myPlot,
	const plot,
	height = 3.0cm,
	width=8.4cm,
	xmin = 1*4*24,
	xmax = 2*4*24,
	ytick = {-2, -1.5, ..., 2.25},
	minor ytick= {-2, -1.75, ..., 2.25},
	yticklabels = {-2, -1.5, ..., 2.25},
	xtick = {96, 120, ..., 192},
	minor xtick= {96, 104, ..., 192},
	xticklabels = {0, 24, ..., 96},
	grid=none,
  	grid style={black,line width=0.025pt,opacity=0.8},
	clip=true,  
    legend columns=3,
    legend style={
      at={(0.4, 1)},
      anchor=south,
      draw=none,
      fill=none,
      legend cell align=left,
      /tikz/every even column/.append style={column sep=0.3cm}
      },
  	title style={
  		at={(0.5, 1.15)},
  		anchor=north,
  	},
	},
	tick label style={font=\small},
	label style={font=\small},
	legend style={font=\footnotesize},
	}

\pgfplotsset{resultsPlotPower/.style={%
    resultsPlot,
    stack plots=y, stack negative=on previous, area style, enlarge x limits=false,
    xlabel = {},
    ymin = -3,
    ymax = 3,
  },
}


\begin{tikzpicture}

\begin{axis}[%
  	resultsPlot,
  	height = 2.2cm,
  	ymin = 0,
  	ymax = 2,
  	ytick = {0, 1, ..., 2},
  	minor ytick= {0, 0.5, ..., 2},
  	yticklabels = {0, 1, ..., 2},
  	legend columns=3,
  	ylabel={$[\unit{MW}]$},
  	xticklabels = {~},
  ]
   
  \addplot [color=colorRes, line legend,stack plots=false,densely dashed] table [x=time, y=d_er, col sep=comma] {./figures/results230320_2.csv};
  \addlegendentry{$d_{\mathrm{e,r}}$}
  \addplot [color=colorDemandHeat, line legend,stack plots=false,densely dashed] table [x=time, y=d_ed, col sep=comma] {./figures/results230320_2.csv};
  \addlegendentry{$d_{\mathrm{e,d}}$} 
  \addplot [color=colorDemandPower, line legend,stack plots=false,densely dashed] coordinates {(0,0) (0,0)};
  \addlegendentry{$\dot{Q}_{\mathrm{d,}2}$}
  	
	\addplot [sharp plot,color=vgOrange, line legend] coordinates {(0,0) (0,0)};
	\addlegendentry{Scenario I} 
	
	\addplot [sharp plot,color=vgDeepBlue, line legend] coordinates {(0,0) (0,0)};
	\addlegendentry{Scenario II} 

\end{axis}

\begin{axis}[shift={(0cm, -1.5cm)},clip=false,
  	resultsPlot,
  	height = 2.5cm,
  	ymin = 0,
  	ymax = 3.1,
  	ytick = {0, 1, ..., 3},
  	minor ytick= {0, 0.5, ..., 3},
  	yticklabels = {0, 1, ..., 3},
  	legend columns=3,
  	ylabel={$[\unit{MW}_{\mathrm{th}}]$},
  	xticklabels = {~},
  ]

  \addplot [color=colorDemandPower, line legend,stack plots=false,densely dashed,clip=false] table [x=time, y=Q_d, col sep=comma] {./figures/results230320_2.csv};
  	
\end{axis}

\begin{axis}[shift={(0cm, -2.5cm)},clip=false,
  	resultsPlot,
  	height = 2cm,
  	ymin = 0.155,
  	ymax = 0.175,
  	ytick = {0.155, 0.175},
  	minor ytick= {0.155, 0.165, 0.175},
  	yticklabels = {0.155, 0.175},
  	legend columns=4,
  	ylabel={$u_{\mathrm{e,t}}\,[\unit{MW}]$},
  	xticklabels = {~},
  ]
  \addplot [color=vgDeepBlue, line legend,stack plots=false,] table [x=time, y=u_et, col sep=comma] {./figures/results230320_2.csv};
  \addplot [color=vgOrange, line legend,stack plots=false] table [x=time, y=u_et, col sep=comma] {./figures/results230320_1.csv};

\end{axis}

\begin{axis}[shift={(0cm, -4.4cm)},clip=false,
  	resultsPlot,
  	height = 2.8cm,
  	ymin = -1.2,
  	ymax = 1.2,
  	ytick = {-1.2, -0.6, 0, 0.6, 1.2},
  	minor ytick= {-1.2, -0.6, 0, 0.6, 1.2},
  	yticklabels = {-1.2, -0.6, 0, 0.6, 1.2},
  	legend columns=4,
  	ylabel={$u_{\mathrm{e,s}}\,[\unit{MW}]$},
  	xticklabels = {~},
  ]
	\PlotLimit{-1.2}
	\PlotLimit{1.2}
  \addplot [color=vgDeepBlue, line legend,stack plots=false,] table [x=time, y=u_es, col sep=comma] {./figures/results230320_2.csv};
  \addplot [color=vgOrange, line legend,stack plots=false] table [x=time, y=u_es, col sep=comma] {./figures/results230320_1.csv};
\end{axis}

\begin{axis}[shift={(0cm, -5.9cm)},clip=false,
  	resultsPlot,
  	height = 2.4cm,
  	ymin = 0,
  	ymax = 1,
  	ytick = {0, 0.5, 1},
  	minor ytick= {0., 0.25, 0.5, 0.75, 1.},
  	yticklabels = {0, 0.5, 1},
  	legend columns=4,
  	ylabel={$u_{\mathrm{e,ph}}\,[\unit{MW}]$},
  	xticklabels = {~},
	legend style={
      	at={(0.6, 0.75)},
      	draw=none,
      	fill=none,
      	/tikz/every even column/.append style={column sep=0.1cm}
      },
  ]
  	\PlotLimit{0}
  	\PlotLimit{1}
  	\addplot [color=vgDeepBlue, line legend,stack plots=false,] table [x=time, y=u_eph, col sep=comma] {./figures/results230320_2.csv};
  	\addplot [color=vgOrange, line legend,stack plots=false] table [x=time, y=u_eph, col sep=comma] {./figures/results230320_1.csv}; 

\end{axis}

\begin{axis}[shift={(0cm, -7.7cm)},clip=false,
  	resultsPlot,
  	height = 2.8cm,
  	ymin = 0,
  	ymax = 5.12,
  	ytick = {0, 1.25, ..., 5},
  	minor ytick= {0, 1.25, ..., 5},
  	yticklabels = {0, 1.25, ..., 5},
  	legend columns=4,
  	ylabel={$x_{\mathrm{e}}\,[\unit{MWh}]$},
  	xticklabels = {~},
	]

  \PlotLimit{0.0}
  \PlotLimit{5.0}
  \addplot[color = vgDeepBlue,] table [x=time, y=x_e, col sep=comma]{./figures/results230320_2.csv};
  \addplot[color = vgOrange] table [x=time, y=x_e, col sep=comma]{./figures/results230320_1.csv};

\end{axis}

\begin{axis}[shift={(0cm, -9.2cm)},clip=false,
  	resultsPlot,
  	height = 2.5cm,
  	ymin = 85,
  	ymax = 95,
  	ytick = {85, 87.5,..., 95},
  	minor ytick= {85, 87.5,..., 95},
  	yticklabels = {85, 87.5,..., 95},
	ylabel = {$T_{\mathrm{e,}1}\,[\unit{^{\circ}C}]$},
	xlabel = {Time step $k\;[1/4\,\unit{h}]$},
	legend columns=2,
	legend style={
      at={(0.5, 0.95)},
      anchor=south,
      draw=none,
      fill=none,
      legend cell align=left,
      /tikz/every even column/.append style={column sep=0.3cm}
      },
	]
,
  \PlotLimit{85.5}
  \PlotLimit{95}
  \addplot[color = vgDeepBlue] table [x=time, y=T_e1, col sep=comma]{./figures/results230320_2.csv};
  \addplot[color = vgOrange] table [x=time, y=T_e1, col sep=comma]{./figures/results230320_1.csv};
\end{axis}

\end{tikzpicture}
	\caption{Assumed load case, optimized control inputs $u_{\mathrm{e}}(k)$, stored electrical energy $x_{\mathrm{e}}(k)$ and supply temperature $T_{\mathrm{e,}1}(k)$ for both scenarios.
	Admissible operating boundaries are indicated by dash-dotted lines.}
	\label{fig:results_unified}
\end{figure}
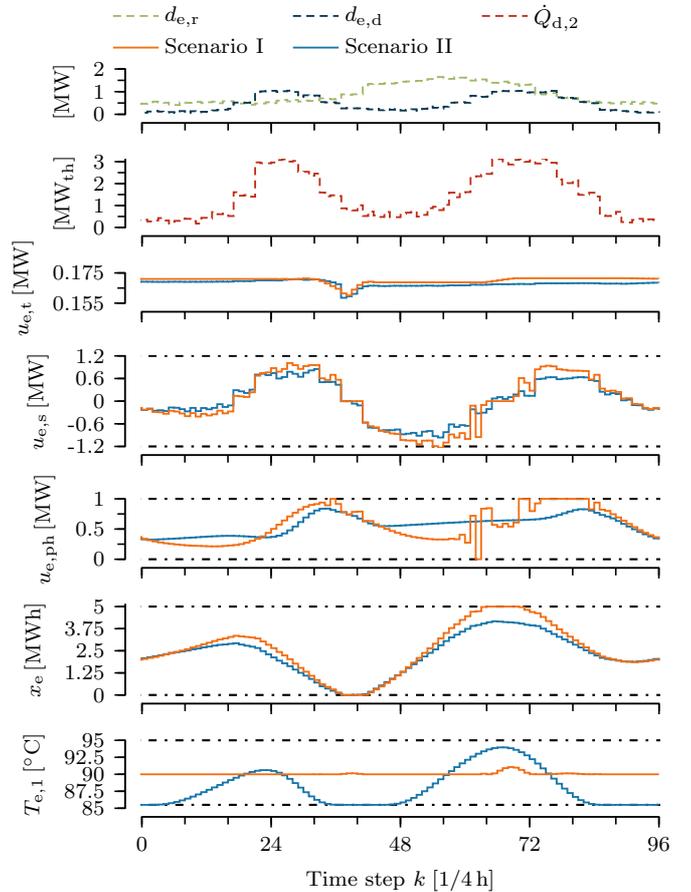

The considered load profiles for $d_{\mathrm{e,r}}(k)$, $d_{\mathrm{e,d}}(k)$, and $\dot{Q}_{\mathrm{d,}2}(k)$,
are shown in the upper plots of Fig.~\ref{fig:results_unified}.
The loads $d_{\mathrm{e,d}}(k)$ and $\dot{Q}_{\mathrm{d,}2}(k)$ have two characteristic peaks that qualitatively correspond to typical household load profiles \citep{meier_Lastprofile_1999}.
The RES infeed $d_{\mathrm{e,d}}(k)$ is based on solar power supply derived from a typical solar radiation curve \citep[see e.g.][]{fan_impacts_2018}.
To be able to exploit characteristic daily patterns, we chose $N=96$ and $\Delta t=15\unit{min}$ ($N \cdot \Delta t = 24\unit{h}$).

The first scenario represents the conventional supply temperature-based operation of DHGs.
That is, MPC~I aims to keep the temperature $T_{\mathrm{e,}1}^{\mathrm{I}}(k)$ close to $x_{\mathrm{t,mp}}^{\mathrm{d}}$, while minimizing $u_{\mathrm{e,t}}^{I}(k)$.
Therefore, we set $x_{\mathrm{t,mp}}(k)=T_{\mathrm{e,}1}^{\mathrm{I}}(k)$ as manipulated temperature and choose the desired temperature $x_{\mathrm{t,mp}}^{\mathrm{d}}=T_{\mathrm{N}}$.
The operating costs $J$  implemented by choosing $c_{\mathrm{e,t}}^{\mathrm{I}}=c_{\mathrm{t}}^{\mathrm{I}}=10$, while the remaining weights are set to zero.
Furthermore, we set $\underline{x}_{\mathrm{t,}l}^{\mathrm{I}}=T_{\mathrm{N}}$ for the nodes or edges $l$ that receive heated water. 

In scenario II, we aim to increase the operational flexibility of the ETMG by utilizing the thermal dynamics.
For this, variable flow temperatures are allowed, i.e., $\mathcal{X}_{\mathrm{t}}=\emptyset$, and economical as well as efficiency-based costs are minimized.
Thus, we set $\underline{x}_{\mathrm{t,}l}^{\mathrm{II}}=85.5^{\circ}\mathrm{C}$ for the nodes or edges $l$ that receive heated water, which corresponds to a permissible deviation of $-5\%$ from $90^{\circ}\mathrm{C}$, and choose $c_{\mathrm{e,s}}^{\mathrm{II}}=c_{\mathrm{e,ph}}^{\mathrm{II}}=0.01$, $c_{\mathrm{e,ph,I}}^{\mathrm{II}}=c_{\mathrm{e,ph,II}}^{\mathrm{II}}=0.1$, $c_{\mathrm{e,t}}^{\mathrm{II}}=10$ and $c_{\mathrm{t}}^{\mathrm{II}}=0$. 

For both scenarios, the lower temperature limits of nodes and edges $l$ containing cooled water as well as the upper temperature limits of all nodes and edges $m$ are $\underline{x}_{\mathrm{t,}l}=55^{\circ}\mathrm{C}$ and $\overline{x}_{\mathrm{t,}m}=95^{\circ}\mathrm{C}$, respectively.
The overall power line and control boundaries are $\overline{p}_{\mathrm{e,}l}=-\underline{p}_{\mathrm{e,}l}=1.2\,\unit{MW}$ for $l=[1,6]\subset\mathbb{N}$, $\overline{u}_{\mathrm{e,}i}=-\underline{u}_{\mathrm{e,}i}=1.2\,\unit{MW}$ for $i\in\{\mathrm{t,\,s}\}$ as well as $\overline{u}_{\mathrm{e,hp}}=0\,\unit{MW}$ and $\underline{u}_{\mathrm{e,hp}}=-1\,\unit{MW}$.
The capacity of the ESS is $\overline{x}_{\mathrm{e}}=5\,\unit{MWh}$.
The heat pump has a nominal power of $u_{\mathrm{hp}}^{\mathrm{d}}=-0.56\,\unit{MW}$ with constant $\alpha=3$, see \eqref{eq:PowerToHeatUnit}.
As stated in Section \ref{subsec:OptimizationModel}, we assume perfect forecast.

By applying MPC~II, the power demand from the main grid is reduced by $1.23\%$ compared to MPC~I, such that $u_{\mathrm{e,t}}^{II}(k)\leq u_{\mathrm{e,t}}^{I}(k)$ for all ${k\in ([0,\ldots,96]\setminus[33,\ldots,37]) \subset\mathbb{N}}$.
Furthermore, the peak powers of the ESS and the heat pump decrease by $21.90\%$ and $16.10\%$, respectively, compared to Scenario~I.
This also results in a reduction of the used ESS capacity by $16.65\%$.
In addition, a uniform operation of the heat pump is achieved, which is clearly shown by a significant decrease in the variance of $u_{\mathrm{e,hp}}^{II}(k)$ by $32.16\%$, compared to the one of $u_{\mathrm{e,hp}}^{I}(k)$.

From the bottom plot of Fig.~\ref{fig:results_unified}, it can be seen that MPC~I successfully maintains $T_{\mathrm{e,}1}^{I}\approx x_{\mathrm{t,mp}}^{\mathrm{d}}$, whereas MPC II utilizes the full range of admissible temperature levels.
MPC~II furthermore enables a preheating of the supply pipe, i.e., an increase in $T_{\mathrm{e,}1}^{II}(k)$, before the thermal demand $\dot{Q}_{\mathrm{d,}2}(k)$ rises (for $k<17$).
This illustrates how the use of a dynamic DHG model in combination with the predictive nature of MPC can anticipate future peak loads.
Storage can be both, dedicated devices, such as the ESS, and inherent storage capacities, such as the water in the DHG pipes.
Finally, the economical and efficiency-based costs of Scenario~I are $18.84\%$ less than those of Scenario~I.

To summarize, the presented case study demonstrates that, by admitting variable flow temperatures, the proposed MPC~II can exploit the inherent thermal storage capacities of the DHG to reduce the 
economic and efficiency-related costs.
Moreover, it decreases the power demand from the main grid and keeps the heat pump operation closer to its optimal operating power, while simultaneously requiring a lower ESS capacity.

\section{Conclusion}\label{sec:Conclusion}
In this paper, we have addressed the necessity of a flexible operation of ETMGs in future climate-neutral energy systems.
More precisely, we have developed a discrete-time state model of an ETMG as well as an MPC scheme that explicitly considers the modeled flow temperature and storage unit dynamics.
Via a numerical case study we have shown that by applying MPC, the operating flexibility of an ETMG can be increased by load shifting and peak shaving. 
This supports an efficient sector-coupled operation of ETMG predominantly demanding RES-based electrical power for both thermal and electrical loads. 

In future work, we plan to further investigate the scalability of the presented approach and seek to incorporate load and generation uncertainties in the control design.

\bibliography{mpc_for_eten}

\end{document}